\documentclass[prc,showpacs,floatfix]{revtex4}
\usepackage{bm}
\usepackage{dcolumn}
\usepackage{graphicx}
\usepackage{amsmath}
\newcommand{\beq}{\begin{equation}}
\newcommand{\eeq}{\end{equation}}
\newcommand{\beqa}{\begin{eqnarray}}
\newcommand{\eeqa}{\end{eqnarray}}
\begin{document}
\title{
{\bf Coherent $\pi^0\eta$ photoproduction on $s$-shell nuclei}}
\author{
M.~Egorov and A.~Fix}%
\affiliation{Laboratory of Mathematical Physics, Tomsk Polytechnic University, Tomsk,
Russia}
\date{\today}
\begin{abstract}
Coherent photoproduction of $\pi^0\eta$ on the deuteron, $^3$He and $^4$He nuclei is
considered in the energy region from threshold to the lab photon energy $E_\gamma=1.2$
GeV. The transition amplitude is derived using impulse approximation. Effects of pion
absorption are included by means of the Fernbach-Serber-Taylor model. Interaction of the
produced $\eta$ mesons with the recoiled nucleus is taken into account for the reactions
on $d$ and $^3$He. The corresponding $\eta d$ and $\eta ^3$He scattering amplitudes are
obtained as solutions of the few-body equations for $\eta NN$ and $\eta-3N$ systems.
Impact of this interaction on the differential cross section in the region of small
relative $\eta$-nuclear momenta is discussed.
\end{abstract}

 \maketitle


\section{Introduction}
Among different photoproduction channels special interest is focused today on the
processes with two pseudoscalar mesons in the final state. Experimental study of $\pi\pi$
and $\pi\eta$ photoproduction on nucleons and nuclei have become an important part of the
research programs of the European Laboratory for Structural Assessment (ELSA) and Mainz
Microtron (MAMI) facilities \cite{KruschePiEta}. In particular, the database for
$\pi^0\eta$ photoproduction was greatly extended by the new very precise measurements,
covering large region of the lab photon energy from threshold up to 3 GeV
\cite{LNS,Horn1,Ajaka,KashevarovEPJ}. Furthermore, a considerable amount of new data for
polarization observables was reported in Refs.\,\cite{Gutz1,Gutz2,Gutz3,KashevarovPLB}.

New experimental results generate a revival of theoretical interest to $\pi^0\eta$
photoproduction. Besides the most early studies of \cite{Doring} a recent, rather
detailed investigation of this process on a free proton was performed in
\cite{Horn2,FKLO,Doring2}. Most of the efforts are directed towards understanding of the
general dynamical properties of those $N^*$ and $\Delta$ resonances which are not very
well seen in the reactions with a single meson and for which only a weak evidence exists
\cite{PDG}. Analysis of the existing data within different models has provided further
insight into the details of the nucleon excitation spectrum, in particular, in the third
and the fourth resonance region.

It is however clear, that systematic study of meson photoproduction requires detailed
information on the same process in nuclei. Here coherent reactions are of special use.
Different works clearly demonstrate their importance, especially in those cases when the
production proceeds dominantly via resonance excitations. One of the main motivations for
studying these reactions is to obtain information on the isotopic structure of the
elementary production amplitude. Evident advantage of using light nuclei as targets is
the small number of nucleons. This allows one to minimize the influence of a nuclear
environment on the elementary process, on the one hand, and to adopt an accurate
microscopic description of the nuclear states, on the other hand.

An important question related to $\pi^0\eta$ photoproduction on nuclei concerns
$\eta$-nuclear interaction in the final state. Although the $\eta$-nuclear scattering
problem is by itself rather many-sided, the major part of the related questions are
connected to the one central point -- $\eta N$ phenomenology in the low-energy regime.
More specifically, the matter concerns determination of the $\eta N$ low-energy
interaction parameters, primarily the scattering length $a_{\eta N}$. In the absence of
the direct scattering results, final state interaction (FSI) in $\eta$ production on
nuclei remains the major source of information on the $\eta N$ dynamics. Typical method
of studying $\eta N$ system in these reactions follows the scheme: (i) $\eta$ production
on nuclei, (ii) $\eta A$ model, (iii) $\eta N$ interaction parameters. The central point
in this sequence is the $\eta$-nuclear interaction model. Since in general one intends to
connect $\eta A$ properties with those of $\eta N$, the $\eta A$ model should be based on
a refined microscopic approach, wherever possible, and at the same time, it should allow
one to take systematically into account fundamental properties of $\eta A$ system, such
as unitarity of the scattering matrix, which is especially important at low energies.
Here the few-body calculations \cite{Shevchenko1,Deloff,Pena1,Wycech,FiArNP,FiAr3b},
utilizing, as a rule, separable $\eta N$ matrix, have already deserved a reputation of an
effective theoretical method. Since these calculations are mostly restricted to the
systems with three and four particles, by now only $\eta$ production on deuterium
\cite{Shevchenko2,Pena2,FiArPLB,FiArDo} and $^3$He
\cite{Wycech3,Shevchenko3,FiAr2,Upadhyay} is considered in detail. According to the
results of Refs.~\cite{Wycech3,FiAr3b,FiAr2}, for 'reasonable' values of $\eta N$
scattering length with $Re\,a_{\eta N}=0.6\pm 0.2$ fm and $Im\,a_{\eta N}=0.3\pm 0.1$ fm,
attraction strength is insufficient to generate bound states of $\eta$ with two- and
three-body nuclei, so that only virtual poles in these systems appear. On the other hand,
for higher values of $Re a_{\eta N}$ about 0.8 fm, the $\eta$-nuclear forces become
nearly strong enough to bind the system. As a consequence, the corresponding virtual pole
lies very close to the physical region, resulting in strong enhancement of the $\eta$
production cross section.

The case of $^4$He is less clear. Firstly, the existing data \cite{Frasc} for the total
cross section of $dd\to\eta^4$He shows no threshold enhancement due to final state
interaction. Furthermore, in the resent experiment of \cite{Adlarson} no signal from the
decay of a hypothetical $\eta$-mesic $^4$He into $\pi^-p\,^3$He in the same reaction
$dd\to\eta^4$He was detected. These results are rather surprising irrespective of
existence of $\eta ^4$He bound states. They mean that the $s$-matrix pole which in the
case of $\eta ^3$He seems to be close to the threshold energy on the Riemann surface,
moves far away from this point when we turn to $^4$He. Even if one takes into account a
larger number of nucleons in $^4$He and profound increase of its density, total
disappearance of a signal from $\eta ^4$He interaction in the measured spectrum is rather
unexpected. Secondly, there are still no correct few-body results for $\eta^4$He due to
difficulty of the corresponding calculation. Less sophisticated theories, like optical
model, are unable to take correctly into account important features of the low-energy
$\eta$-nuclear interaction, for example, importance of the virtual target excitations
between the successive scattering acts (see, e.g., \cite{FiAr2}). Therefore, even the
qualitative results obtained within this approach are unreliable. The matter is further
complicated by the fact that it is the $\eta ^4$He case, where the binding may appear.
The value of the scattering length corresponding to a weakly bound or virtual state is
known to be very sensitive to small variation of the potential parameters. Therefore, it
may turn out that if we apply few-body formalism to $\eta^4$He, the result will strongly
depend not only on the $\eta N$ parameters, but also on the approximations used for
$^4$He states (e.g., on details of the $NN$-forces at short distances, inclusion of the
repulsive core into $NN$ potential, $\it etc.$). This will lead to strong model
dependence of the calculation, not to mention that the five-body scattering problem is
technically very difficult by itself.

Since no microscopic $\eta A$ calculations are available for the nuclei with $A>3$, the
methods allowing model independent extraction of the $\eta A$ scattering parameters
directly from the measured observables are of special importance. Several steps are
already done in
Refs.\,\cite{Germond,Calen1,Bilger,Smyrski,Adam,Mersman,Pheron,Krzemien,Baskov,Wilkin,Niskanen},
where information on the $\eta$-nuclear scattering from the characteristic behavior or
the cross section in the region of small $\eta A$ relative energies was obtained. As a
rule, the underlying method is based on the approach developed by Watson \cite{Watson}
and Migdal \cite{Migdal}. Under certain conditions this approach makes it possible to
study $\eta A$ interaction directly from a distribution over the relative $\eta A$
energy. The case in point is a characteristic enhancement of the $\eta$ yield due to
strong attraction between $\eta$ meson and the recoiled nucleus. Here again, coherent
photoproduction of $\pi^0\eta$ seems to have some advantages over other reactions, like
$A(p,\eta p)A$ or $A(\pi,\eta N)B$, for which one has to take into account interaction
between all three final particles. Even the simplest case, when the composite nature of
the nucleus is ignored (e.g., if one neglects its virtual excitations), requires
three-body calculation. This problem should be much less important for $\pi^0\eta$
photoproduction. In this case one can safely neglect the interaction between $\eta$ and
$\pi^0$, since in the energy region under discussion $E_\gamma\leq 1.5$\,GeV this system
does not resonate. As a result, the whole interaction process may be approximated by the
sum of $\eta A$ and $\pi^0 A$ rescatterings.

Although the reactions $A(\gamma,\pi^0\eta)A$ offer important advantages, up to now
rather little effort has been devoted to their theoretical study. Perhaps, the major
reason of this fact is
difficulty of the experimental identification of these reactions
due to smallness of the coherent cross section in comparison to the background quasifree
process $A(\gamma,\pi^0\eta N)$. This situation should change with new measurements of
$\pi^0\eta$ photoproduction on nuclei \cite{KruschePiEta}. In anticipation of the new
data, we present here theoretical results for $\pi^0\eta$ photoproduction on $s$-shell
nuclei, for which we have chosen as specific examples $d$, $^3$He, and $^4$He.

\section{Model ingredients}\label{Sect2}

We consider the process
\begin{equation}\label{20}
\gamma(E_\gamma,\vec{k};\,\vec{\varepsilon}_\lambda) +
A(E_A,\vec{Q}_A)\to\pi^0(\omega_\pi,\,\vec{q}_\pi) + \eta(\omega_\eta,\,\vec{q}_\eta)
+A(E_A^\prime,\vec{Q}_A^\prime)\,,
\end{equation}
where the 4-momenta of the participating particles are given in the parentheses. The
calculations are performed in the laboratory frame where the initial nucleus $A$ is at
rest ($Q_A=0$). The circular polarization vector of the photon is denoted by
$\vec{\varepsilon}_\lambda$ with $\lambda=\pm 1$. One of the features of the reaction
(\ref{20}) is rather high momentum transfer associated with a relatively large mass of
the $\pi\eta$ system. This firstly results in rather low cross section (several hundreds
nb) and, moreover, in sensitivity of its magnitude to details of the nuclear wave
function at short internuclear distances. At the same time, this sensitivity does not
necessarily mean that a refined microscopic nuclear model is required for the
calculation. Indeed, according to our general notion about the coherent production, its
basic mechanism, yielding the main fraction of the amplitude, is impulse approximation
accompanied by the final state interaction (FSI) effects. Within this model, the
unpolarized cross section is mainly governed by the nuclear form factor. The latter is
free from the ambiguities of the nuclear structure and may well be described
phenomenologically without resorting to complicated microscopic calculations.
Furthermore, interaction of pions with nuclei in the resonance region, where its main
effect is attenuation of the pion wave function inside the nucleus, may be described in
terms of the pion mean free path in the nuclear matter. As for the $\eta$-nuclear
interaction, which is one of the main objects of the present study, in the low energy
region it is mainly determined by the long-range part of the $\eta$-nuclear wave
function, and therefore, should be insensitive to the structural details of the nuclear
model. Taking account of this observation,
we use for the nuclear wave functions the phenomenological models, which reproduce the
corresponding formfactors up to the values of momentum transfer, which are characteristic
for $\pi\eta$ production in the second and the third resonance region. For the deuteron
we take the wave function of the Bonn potential (full model)~\cite{Machleidt}. For $^3$He
target the separable parametrization from \cite{JulichWF} is employed, and for $^4$He we
adopt Fourier transform of the $r$-space wave function from \cite{Sher83}.

\begin{figure}
\begin{center}
\resizebox{0.8\textwidth}{!}{%
\includegraphics{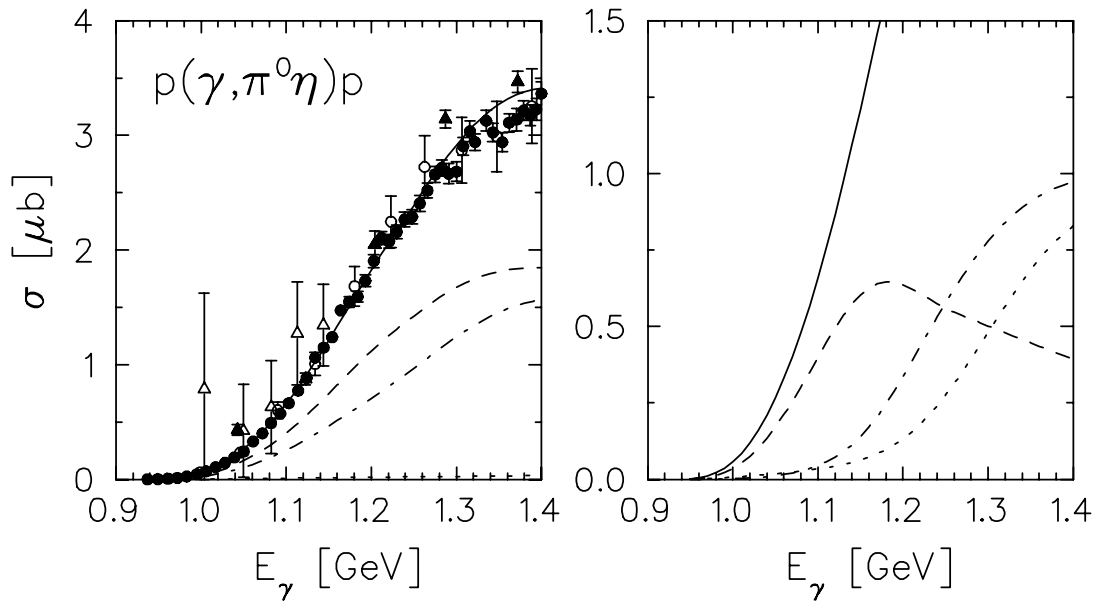}}
\caption{Total cross section for $\gamma p\to\pi^0\eta p$ calculated using the isobar
model of Ref.\,\cite{FKLO} (solid curve on both panels). On the left panel, the dash-dotted
and the dashed curves correspond to the spin-flip part $\vec{K}$ and the spin
independent part $L$ in Eq.~(\protect\ref{30}). The dotted curve shows the contribution
of the isoscalar part $A^{(0)}$ of the amplitude (\ref{Isospin}). The data are from
Ref.\,\cite{LNS} (empty triangles), \cite{Ajaka} (empty circles), \cite{Horn1} (filled
triangles), and \cite{KashevarovEPJ} (filled circles). On the right panel, the dashed and
dash-dotted curves show the contribution from the resonances $D_{33}(1700)$ and
$D_{33}(1940)$, respectively. The dotted curve is the combined contribution of the
remaining resonances ($P_{33}(1600)$, $P_{31}(1750)$, $F_{35}(1905)$, $P_{33}(1920)$) and
the Born terms.} \label{fig1}
\end{center}
\end{figure}

The elementary operator has the well known form
\begin{equation}\label{30}
t_{\gamma N\to \pi\eta N}=L+i\vec{K}\cdot\vec{\sigma}\,,
\end{equation}
reflecting the general spin structure of photoproduction of pseudoscalar mesons on spin
1/2 fermions. Here we used the isobar model from Ref.\,\cite{FKLO} (first solution).
Apart from the dominant $D_{33}(1700)$ and $D_{33}(1940)$ resonances, the model
\cite{FKLO} also contains the positive parity states $P_{33}(1600)$, $P_{31}(1750)$,
$F_{35}(1905)$, $P_{33}(1920)$ and relatively small admixture of the Born terms. The
amplitude is diagrammatically presented in Fig.\,2 of Ref.\,\cite{FKLO}. The parameters
were fitted to the experimental angular distributions of pions and $\eta$ mesons in
$\gamma p\to\pi^0\eta p$.
In Fig.\,\ref{fig1} we show the total cross section of this reaction. The contributions
of different terms are separately presented on the right panel. According to the
calculation, in the region up to $E_{\gamma}=1.2$\,GeV the major fraction of the cross
section is provided by the resonance $D_{33}(1700)$.

On the left panel of Fig.\,\ref{fig1} we also plotted the components $\sigma_K$ and
$\sigma_L$ of the cross section coming from the spin-flip and spin independent part of
the operator (\ref{30}). Neglecting all terms apart from $D_{33}(1700)$ one can obtain
for the ratio of $\sigma_K$ to $\sigma_L$
\begin{equation}\label{ratKL}
\frac{\sigma_K}{\sigma_L}=\frac12+\frac12\left(\frac{3-\sqrt{3}a}{1+\sqrt{3}a}
\right)^2\,,
\end{equation}
where $a$ is the ratio of $3/2$ to $1/2$ helicity amplitudes of the $D_{33}(1700)$
resonance
\begin{equation}
a=A_{3/2}/A_{1/2}\,.
\end{equation}
In the model \cite{FKLO} this parameter changes from 0.9 to 1.1 in the region
$E_\gamma\leq 1.2$\,GeV, where $D_{33}$(1700) dominates (see Fig.\,6 in \cite{FKLO}).
Therefore, at these photon energies the ratio (\ref{ratKL}) remains almost constant and
is equal to $0.60\pm 0.04$, so that the components $\sigma_K$ and $\sigma_L$ are
comparable, as may also be seen from Fig.\,\ref{fig1}.

Within the impulse approximation the amplitude on a nucleus is proportional to that on a
single nucleon sandwiched between the states including the initial and the final nucleus
in the ground state with the spin $J$
\begin{equation}\label{40}
T_{\lambda MM'}=A\langle \vec{q}_\eta,\vec{q}_\pi;JM'|t_{\gamma N\to \pi\eta
N}(\omega)|\vec{k},\lambda;JM\rangle\,,
\end{equation}
with $A$ denoting the number of nucleons in the target. The quantity $\omega$ in
(\ref{40}) has a meaning of the invariant $\gamma N$ energy when the nucleon is on the
mass shell. To take properly into account the Fermi motion effect, which should be
important in the resonance region, we use the prescription \cite{Lazard,Dover}, in which
the elementary operator is frozen at the average effective nucleon momentum in the
laboratory system
\begin{equation}\label{rec}
\vec{p}_i=\langle \vec{p}_i\rangle=-\frac{A-1}{2A}\,\vec{Q}\,,
\end{equation}
where $\vec{Q}=\vec{k}-\vec{q}_\pi-\vec{q}_\eta$ is the momentum transferred to the
nucleus and $A$ is the nuclear mass number. This choice is compatible with the
requirements of energy and momentum conservation together with the on-mass-shell
conditions for the nucleon both in the initial and the final state.

The unpolarized cross section of the reaction (\ref{20})
is proportional to the square of the amplitude (\ref{40}) averaged over the spin states
\begin{equation}\label{50}
\frac{d\sigma}{d\Omega_\pi d\omega_\pi d\Omega_{\eta}^*}=
\frac{1}{(2\pi)^5}\frac{E'_Aq_\pi q_{\eta}^*}{8E_\gamma\omega_{\eta A} W}
\frac{1}{2(2J+1)}\sum\limits_{\lambda MM'}|T_{\lambda MM'}|^2\,,
\end{equation}
where the total $\eta A$ energy $\omega_{\eta A}$, as well as the $\eta$ momentum
$\vec{q}^{\,*}_\eta$ and the corresponding solid angle $\Omega_\eta^*$ are calculated in
the $\eta A$ center-of-mass (c.m.) frame.

Of primary importance for the coherent reaction is a relative contribution of the
transitions with isospins $I=1/2$ and $I=3/2$. Since $\eta$ is an isoscalar particle, the
isospin structure of the operator (\ref{30}) is similar to that of single pion
photoproduction. In particular, for $\pi^0\eta$ we have
\begin{equation}\label{Isospin}
t_{\gamma N\to \pi\eta N}=A^{(0)}\tau_3+\frac{1}{3}A^{(1/2)}+\frac{2}{3}A^{(3/2)}\,,
\end{equation}
where $\tau_3$ is the third component of the nucleon isospin operator $\vec{\tau}$. The
amplitudes $A^{(1/2)}$ and $A^{(3/2)}$ in (\ref{Isospin}) are related to the final
$\pi^0\eta N$ state with the isospin $I=1/2$ and $I=3/2$, respectively. The isoscalar
amplitude $A^{(0)}$ leads only to the states with isospin $1/2$. According to the
analyses of \cite{Doring,Horn2,FKLO} in the region $E_\gamma=1-1.2$ GeV the process
(\ref{30}) is dominated by excitation of $\Delta$-like resonances, so that the role of
the $A^{(0)}$ and $A^{(1/2)}$ components is small in the energy region considered.
Therefore, the resulting elementary cross section is practically the same for proton and
neutron targets. For the reactions on nuclei this means that the effect of coherence is
maximal and the cross section does not depend on the isospin of the target.

\begin{figure}
\begin{center}
\resizebox{0.98\textwidth}{!}{%
\includegraphics{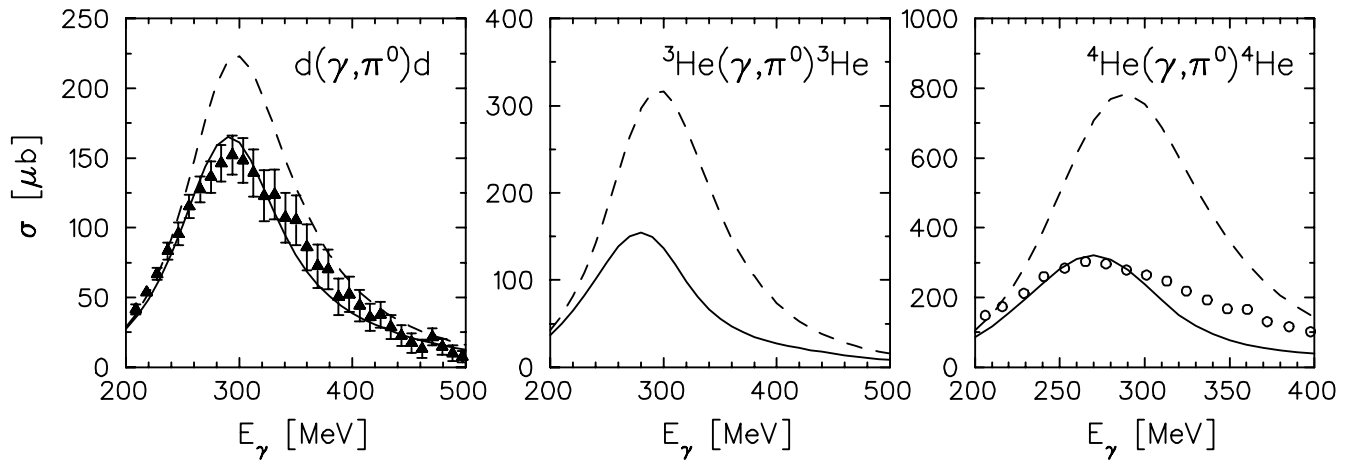}}
\caption{Total cross section for coherent $\pi^0$ photoproduction on $d$, $^3$He, and
$^4$He. The solid (dashed) curves are calculated using impulse approximation with
(without) pion absorption. The elementary $\gamma N\to\pi^0 N$ amplitude is taken from
the MAID2007 model \cite{MAID2007}. The data are from \cite{Kruschepi0d} (triangles) and
\cite{Kruschepi0He4} (circles).} \label{fig2}
\end{center}
\end{figure}

To take into account interaction between the emitted pion and the final nucleus we used a
simplified model in which this interaction is described in terms of classical propagation
of a pion in a nuclear matter. In the resonance region the major impact of a nucleus is
attenuation of the pion beam due to absorption. Apart from the true absorption on nucleon
pairs, the inelastically scattered pions, which in fact contribute to the incoherent
cross section, are also treated as if they are absorbed. Additional interaction effect,
which however should be less important in the region considered, comes from the
modification of the pion wave number in a nuclear medium, which in particular results in
changing diffraction patterns which are characteristic for coherent pion photoproduction
in the resonance region. Here we neglect the last effect and take into account only
absorption of the produced pions using the simple prescription. Namely, the cross section
is multiplied by the energy dependent damping factor, which was calculated as follows.
The pion wave function inside the nucleus was taken in the form used by Fernbach, Serber,
and Taylor \cite{FST} for neutron interactions in nuclei:
\begin{equation}\label{phiFST}
\phi_{\vec{q}_\pi}^{(-)}(\vec{r}\,)=\exp{\big(-i\vec{q}_\pi\cdot\vec{r}\,\big)}D(\vec{r}\,)\,,\quad
D(\vec{r}\,)=\exp{\big(-l(\vec{r}\,)/2\lambda\big)}\,,
\end{equation}
where the damping factor $D(\vec{r}\,)$ depends on the distance $l(\vec{r}\,)$, measured
along the classical trajectory of a meson between the point where it was produced and the
point where it escaped from the nucleus. The optical properties of the nuclear
environment are determined by the mean free path $\lambda$ of a pion. It can be expressed
in terms of the $\pi N$ scattering cross section $\sigma_{\pi N}$ averaged over protons
and neutrons as
\begin{equation}\label{opt_param}
\lambda=\frac{1}{\rho\sigma_{\pi N}}\,,
\end{equation}
where $\rho$ is the nuclear density.
For
simplicity we take the damping factor $D(\vec{r}\,)$ out of the matrix element at a mean
value
\begin{equation}\label{D}
\overline{D(\vec{r}\,)}=\frac{1}{A}\int_V D(\vec{r}\,)\rho(r)d^3r\,,
\end{equation}
where $V$ is the nuclear volume. For the nuclear density $\rho(r)$ in (\ref{opt_param})
and (\ref{D}) a simple hard sphere form
\begin{eqnarray}\label{rho}
\rho(r)&=&\frac{3A}{4\pi R^3}\,,\quad r<R\,,\\
       &=&\quad 0\,,\quad \quad r>R\,,\nonumber
\end{eqnarray}
was taken, where for $R$ we used the r.m.s.\ radius $R=\sqrt{\langle r^2\rangle}$. Then
the absorption effect results in suppressing the cross section by the factor (see also
\cite{Engelbrecht})
\begin{equation}\label{Engel}
\overline{D(\vec{r}\,)}^{\ 2}=\frac{9}{4x^2}\left\{1-\frac{2}{x^2}\left[1-(1+x)e^{-x}
\right]\right\}^2\,,\quad x=\frac{R}{\lambda}\,.
\end{equation}
In Fig.\,\ref{fig2} we show our results for coherent single $\pi^0$ photoproduction on
all three nuclei in the first resonance region. As one can see, for the reaction on $d$
and $^4$He the method provides the required suppression in the region of maximum at
$E_\gamma\approx 280$\,MeV. Above $E_\gamma\approx 320$\,MeV the calculated cross section
underestimates the experimental results. However, in view of extreme simplicity of our
model the agreement with the data is quite reasonable.

Contrary to the pion case, interaction in the $\eta$-nucleus system, where the major role
is played by the strong $s$-wave attraction, is mostly important at lower relative
energies.
This well known property is also observed in the $pd\to p\eta d$ reaction
\cite{Bilger,Upadhyay} where it leads to a pronounced peak in the distribution over the
relative $\eta$-nuclear energy $E$ in the region of small $E$ values. To extract the
$\eta A$ interaction parameters, one assumes that this FSI effect is independent on the
$\eta$ production mechanism, and, therefore, can be unambiguously isolated. In fact, this
assumption is justified only if the following conditions are fulfilled
\cite{Watson,Migdal}: (i) the driving reaction mechanism ($\pi^0\eta$ photoproduction in
our case) is of short-range nature, i.e. its effective radius is essentially smaller than
the characteristic range of $\eta A$ forces, and (ii) attraction between the particles is
comparatively strong and is characterized by low relative momentum, so that it acts
during a sufficiently long time. For the reactions with more than two strongly
interacting particles in the final state the third obvious condition should be added:
(iii) other particles having high velocities, quickly escape the region in which the
production mechanism works, and thus have little effect on the interacting pair. From the
three conditions above the first two seem to be fairly well satisfied in our case.
Indeed, the smallness of the effective range of an interaction responsible for $\pi\eta$
production is ensured by rather large momentum transfer. Strong $\eta A$ attraction is
due to the nearby pole in the corresponding $s$-wave amplitudes. The third condition is
fulfilled due to the small pion mass, which results in rather high velocity of the
produced pion, so that in the major fraction of the reaction events it is at a distance
well removed from the $\eta A$ pair. This latter aspect is considered in the next section
in more detail.

To take into account the $\eta$-nuclear FSI, we add to the impulse approximation $T^{IA}$
the interaction term $T^{\eta A}$
\begin{equation}\label{70}
T=T^{IA}+T^{\eta A}\,,
\end{equation}
with
\begin{equation}\label{80}
T^{\eta A}=\int T^{IA}(\vec{q}\,')G_{\eta A}(q_{\eta}^*,q')t_{\eta
A}(q_{\eta}^*,q';\theta)\frac{d^3q'}{(2\pi)^3}\,.
\end{equation}
Here, $t_{\eta A}(q_{\eta}^*,q';\theta)$ with $\cos\theta=\hat{q}_{\eta}^*\cdot\hat{q}'$
is the half-off-shell $\eta A$ $t$-matrix. In the on-shell region it is related to the
$\eta A$ scattering amplitude as
\begin{equation}\label{90}
t_{\eta A}(q_{\eta}^*,q_{\eta}^*;\theta)=-\frac{2\pi}{\mu_{\eta A}}f_{\eta
A}(q_{\eta}^*,\theta)
\end{equation}
with $\eta$-nuclear reduced mass denoted by $\mu_{\eta A}$. The function $G_{\eta A}$ in
Eq.~(\ref{80}) stands for the $\eta A$ propagator in the momentum representation
\begin{equation}
G_{\eta A}(q_{\eta}^*,q')=\frac{1}{\omega_{\eta A}-q^{\prime 2}/2\mu_{\eta
A}+i\epsilon}\,.
\end{equation}
Due to the $s$-wave character of the main $\eta N$ interaction mechanism, excitation of
the resonance $S_{11}(1535)$, practically the whole amplitude $f_{\eta A}$ is saturated
by its $s$-wave part $f_0$:
\begin{equation}\label{100}
f_{\eta A}(q^*_\eta,\theta)\approx f_0(q^*_\eta)\,.
\end{equation}
We used three different sets of $\eta N$ scattering matrix parameters which give $\eta A$
scattering lengths and effective ranges listed in Table \ref{ta1}. For orientation also
the corresponding values of $\eta N$ scattering length $a_{\eta N}$ are presented. Since
there is no microscopic calculation for $\eta ^4$He, the $\eta$-nuclear interaction was
taken into account only for the reactions on a deuteron and $^3$He, whereas for $^4$He
only absorption of the produced pions was included.

\begin{table*}
\renewcommand{\arraystretch}{2.0}
\caption{Parameters for $\eta d$ and $\eta^3$He low-energy scattering. The values are
obtained by fitting the scattering amplitudes, calculated within the few-body formalism
described in Refs.~\cite{FiArNP} and \cite{FiAr2}. For orientation the corresponding
values of $\eta N$ scattering lengths are also given.} \label{ta1}
\begin{center}
\begin{tabular*}{17cm}
{c|@{\hspace{0.8cm}}c@{\hspace{0.8cm}}|@{\hspace{0.77cm}}c@{\hspace{0.97cm}}c@{\hspace{0.77cm}}
|@{\hspace{0.77cm}}c@{\hspace{0.97cm}}c@{\hspace{0.77cm}}} \hline\noalign{\smallskip} &
$a_{\eta N}$\,[fm], Ref. & $a_{\eta d}$\,[fm] & $r_{0\,\eta
d}$\,[fm] & $a_{\eta^3\!He}$\,[fm] & $r_{0\,\eta^3\!He}$\,[fm] \\
\noalign{\smallskip}\hline\noalign{\smallskip}
1 & $0.50+i\,0.33$\,, \cite{Wilkin1}  & $1.232+i\,1.110$ & $2.429-i\,1.037$ & $1.866+i\,2.752$ & $1.934-i\,0.532$ \\
2 & $0.75+i\,0.27$\,, \cite{Wyc}  & $2.221+i\,1.153$ & $1.870-i\,0.471$ & $4.199+i\,4.817$ & $1.442-i\,0.139$ \\
3 & $1.03+i\,0.41$\,, \cite{Work}  & $3.318+i\,2.648$ & $1.818-i\,0.456$ & $-3.767+i\,9.362$ & $1.631-i\,0.187$ \\
\noalign{\smallskip}\hline
\end{tabular*}
\end{center}
\end{table*}

\section{Discussion of the results}\label{Sect3}

\begin{figure*}
\begin{center}
\resizebox{1.0\textwidth}{!}{%
\includegraphics{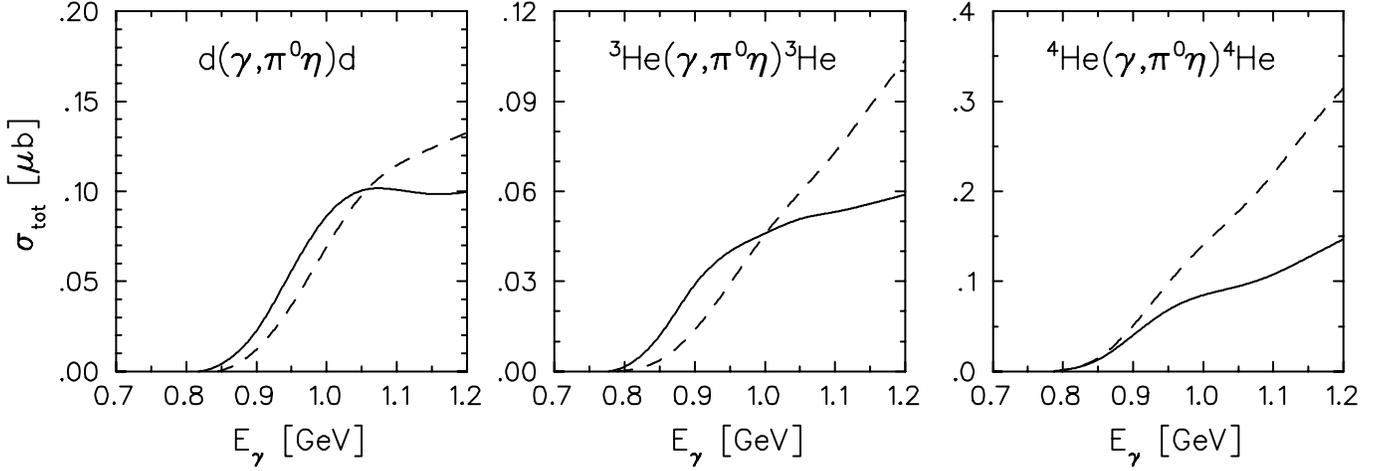}}
\caption{Total cross section for $\pi^0\eta$ photoproduction on the deuteron, $^3$He and
$^4$He. The solid (dashed) curves are obtained with (without) inclusion of interaction in
the final state. For $\eta A$ interaction the set of parameters 2 in Table \ref{ta1} was
used. In the case of $^4$He only pion absorption was taken into account.} \label{fig3}
\end{center}
\end{figure*}

We would like to start our discussion from the total cross section presented in
Fig.\,\ref{fig3}. As noted above, the cross section value is independent of the nuclear
isospin and should be mostly determined by the spin of the target and its density. In
particular, the cross section turns out to be rather sensitive to the details of the
target wave function. For example, if one neglects the deuteron $d$-wave component, the
total cross section for $d(\gamma,\pi^0\eta)d$ is reduced by about 30\,$\%$ at
$E_\gamma=1$\,GeV. Furthermore, since, as discussed above, the contributions of the
spin-flip and spin independent part in $\pi^0\eta$ photoproduction are comparable, the
nuclear cross section strongly depends on the nuclear spin. The interplay between the
nuclear fromfactor at high momentum transfer and the spin structure of the production
matrix element leads to nontrivial dependence of the cross section on the choice of the
target. As we can see from Fig.\,\ref{fig3}, without FSI the deuteron cross section turns
out to be almost twice as large as that on $^3$He.

Since for $\pi$-nucleus interaction we take into account only absorption, the only
influence of $\pi A$ FSI is attenuation of the cross section. Clearly, it should increase
with increasing number of nucleons and increasing the nuclear density. This means that
among the three nuclei considered here the largest effect should be observed for $^4$He.
Furthermore, absorption is known to be especially important in the region of
$\Delta(1232)$, where $\pi A$ scattering becomes highly inelastic, leading for heavier
nuclei to the so-called surface production mechanism. In our case it is responsible for a
significant reduction of the total cross section especially above $E_\gamma=1$\,GeV.

\begin{figure*}
\begin{center}
\resizebox{0.4\textwidth}{!}{%
\includegraphics{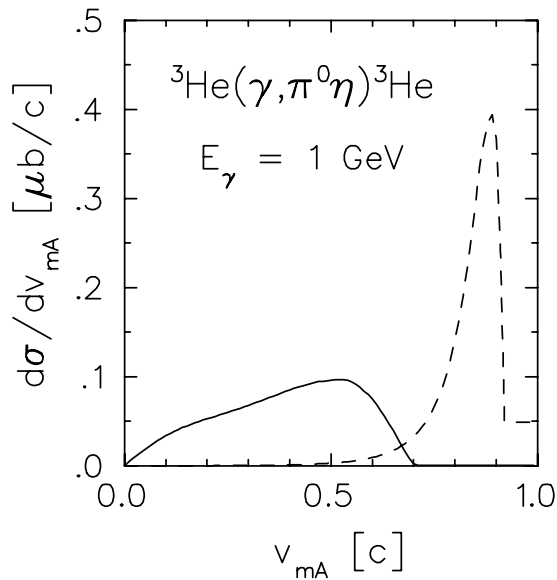}}
\caption{Distribution of the cross section for $^3$He$(\gamma,\pi^0\eta)^3$He over the
relative velocity in the $\eta ^3$He (solid line) and $\pi ^3$He (dashed line)
subsystems.} \label{fig4}
\end{center}
\end{figure*}

As already noted above, an important feature of the reaction (\ref{20}) is that the time
which the produced pion spends in the interaction region is short in comparison to that
for the $\eta$ meson. To demonstrate this feature we present in Fig.\,\ref{fig4} the
distribution of the cross section for $^3$He$(\gamma,\pi^0\eta)^3$He over the relative
velocity in the $\eta ^3$He and $\pi ^3$He subsystems. The velocity was calculated in the
corresponding $\eta A$ and $\pi A$ center-of-mass (c.m.) frames as
\begin{equation}\label{vxA}
v_{mA}=\frac{v^*_m+v^*_A}{1+v^*_mv^*_A}=\frac{\lambda^{1/2}(\omega_{mA}^2,M_m^2,M_A^2)}{\omega_{mA}^2-M_m^2-M_A^2}\,,\quad
m\in\{\pi,\eta\}\,,
\end{equation}
where $v^*_m$ and $v^*_A$ are the c.m.\ velocities of the meson $m$ and the residual
$^3$He nucleus, and the triangle function $\lambda$ is defined as
\begin{equation}
\lambda(x,y,z)=(x-y-z)^2-4yz\,.
\end{equation}
As one can see, the maximum in the distribution over $v_{\pi A}$ is shifted to much
higher values with respect to the maximum of $d\sigma/dv_{\eta A}$. The corresponding
average values of $v_{mA}$ are $v_{\eta A}=0.38$\,c and $v_{\pi A}=0.83$\,c.

Using the distribution in Fig.\,\ref{fig4} one can estimate the characteristic time the
meson $m$ takes to propagate a scattering center of radius $R$ as
\begin{equation}\label{tx}
t_m=\frac{R+1/q^*_m}{v_{mA}}+Q_m\,,
\end{equation}
where the wavelength $1/q^*_m$ takes into account the wave properties of a particle and
$Q_m$ is a time delay due to attraction. Taking $v_{\pi A}=0.83$\,c, $q^*_\pi=190$\,MeV
(corresponds to $v_{\pi A}$), $R=2$\,fm, and $Q_\pi=1/\Gamma_{\Delta}\approx
1/120$\,MeV$^{-1}$ one obtains for the pion $t_\pi\approx 1.62\cdot 10^{-23}$\,s.

To calculate $Q_\eta$ one can use the Eisenbud-Wigner formula \cite{EisWig}
\begin{equation}
Q_\eta=2\frac{d}{dE}\delta_0(E)\,,
\end{equation}
where $\delta_0(E)$ is the phase shift of the $s$-wave $\eta A$ scattering, for which one
can take
\begin{equation}
\delta_0(E)\approx Re\,a_{\eta A}\sqrt{2\mu_{\eta A}E}\,.
\end{equation}
The relative $\eta A$ energy $E$ is determined as
\begin{equation}
E=\omega_{\eta A}-M_\eta-M_A\,,
\end{equation}
where $M_A$ is the mass of the nucleus. Taking $Re\,a_{\eta ^3\!He}=4$\,fm from Table
\ref{ta1} and $E=38$ MeV (corresponds to the average relative velocity $v_{\eta
A}=0.38$\,c), one obtains for the time delay $Q_\eta=6.67\cdot 10^{-23}$\,s, so that the
resulting value of $t_\eta$ (\ref{tx}) turns out to be $9.34\cdot 10^{-23}$\,s, almost
six times larger than $t_\pi$. This result supports our intuitive assumption that the
pion tends to quickly escape the interaction region and its presence should have little
effect on the $\eta A$ interaction.

\begin{figure*}
\begin{center}
\resizebox{0.8\textwidth}{!}{%
\includegraphics{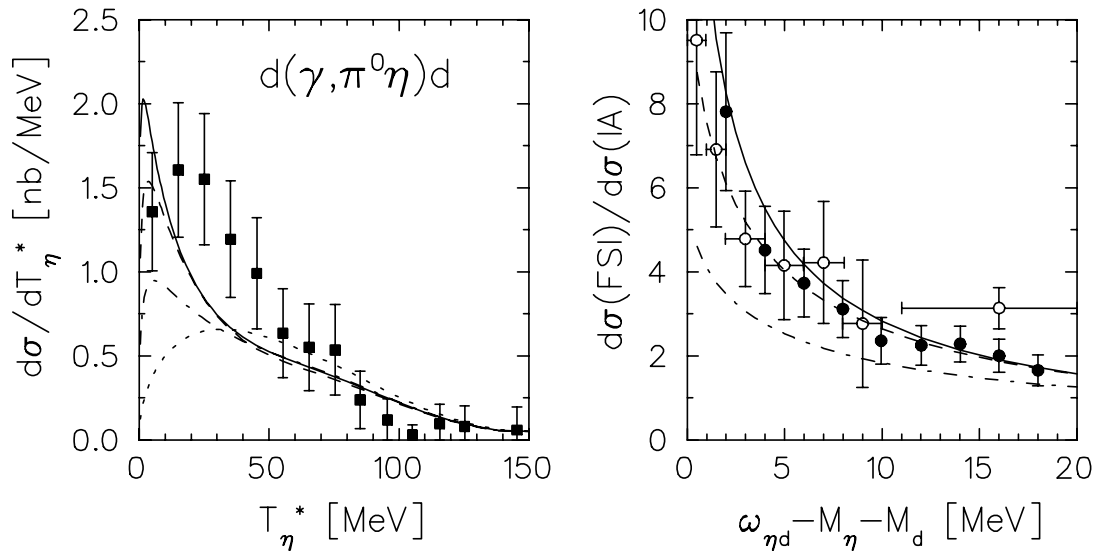}}
\caption{Left: kinetic energy spectrum of $\eta$-mesons in the reaction
$d(\gamma,\pi^0\eta)d$ averaged over the energy range $E_\gamma=0.9-1.1$ GeV. The dotted
curve is calculated without $\eta d$ interaction. The dash-dotted, dashed, and the solid
curves are obtained with the sets 1,2, and 3 of $\eta A$ scattering parameters listed in
Table \ref{ta1}. The filled squares are the preliminary data from
Ref.\,\cite{KruschePiEta}. Right: ratio of the FSI to IA cross sections plotted against
the relative $\eta d$ energy. Notations of the curves as on the left panel. Empty and
filled circles show the $pn\to\eta d$ and $pd\to\eta pd$ cross sections from
Refs.\,\cite{Calen1} and \cite{Bilger}, respectively, divided by the arbitrarily
normalized phase space.} \label{fig5}
\end{center}
\end{figure*}

\begin{figure*}
\begin{center}
\resizebox{0.8\textwidth}{!}{%
\includegraphics{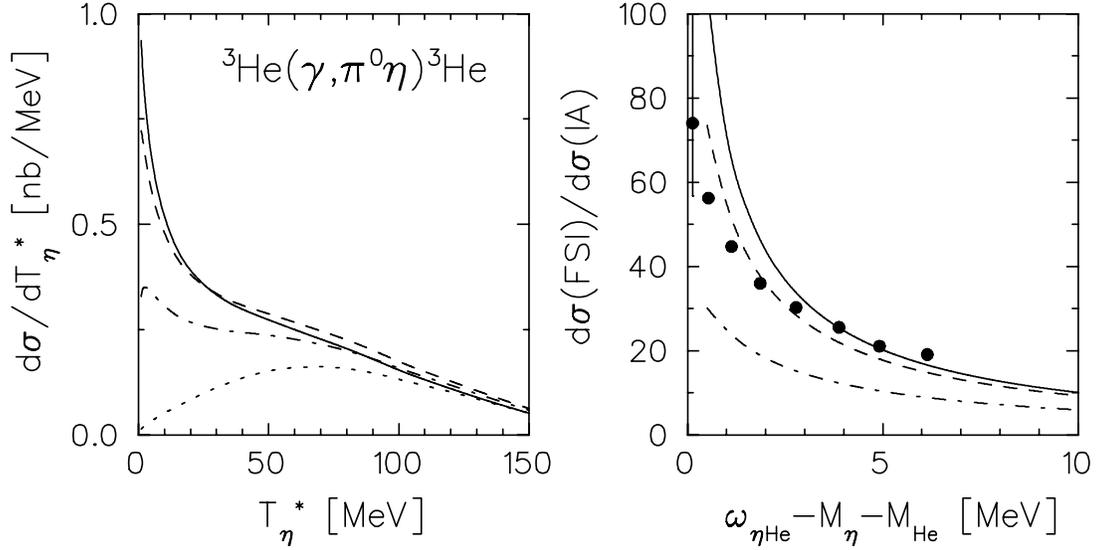}}
\caption{Same as in Fig.\,\ref{fig5} for $^3$He$(\gamma,\pi^0\eta)^3$He. The points on
the right panel show the $pd\to\eta^3$He cross section from Ref.\,\cite{Mayer} divided by
the phase space. The normalization of the data is arbitrary.} \label{fig6}
\end{center}
\end{figure*}

To demonstrate the role of $\eta$-nuclear interaction, we show in Figs.~\ref{fig5} and
\ref{fig6} the distribution over the relative $\eta A$ energy $E$. As expected, the
spectrum rises rapidly from zero and exhibits a peak very close to the lower limit $E=0$.
On the whole, inclusion of $\eta A$ FSI enhances the $\eta$ yield. The resulting total
cross in Fig.\,\ref{fig3} is visibly increased due to $\eta A$ attraction in the region
up to $E_\gamma=1$\,GeV. At higher energies, the pion absorption takes over leading to
the eduction of the cross section.

In the region of low values of the relative $\eta A$ momenta $q^*_\eta R\ll 1$, the shape
of the spectrum close to the peak may be described in a simple manner as
\cite{Watson,Migdal}
\begin{equation}\label{170}
\frac{d\sigma}{dE}(E)\sim P(E)\,|f_0(E)|^2\,,
\end{equation}
where $f_0(E)$ is the $\eta A$ scattering amplitude and $P(E)$ is the reaction phase
space. Using the effective range formula
\begin{equation}\label{130}
q_\eta^*\cot\delta_0=\frac{1}{a}+\frac{r_0}{2}q_\eta^{*2}\,,
\end{equation}
one can expand the ratio $P(E)/\sigma(E)$ in powers of the momentum $q_\eta^*$
\begin{equation}\label{polyn}
\frac{P(E)}{d\sigma/dE}\sim\frac{1}{|f_0(E)|^2}=q_\eta^{*2}\left|\cot\delta_0-i\right|^2=\sum_{n=0}^4C_nq_\eta^{*n}\,.
\end{equation}
Since the scattering length $a_{\eta A}$ has nonzero imaginary part, the expansion
(\ref{polyn}) contains odd powers of $q_\eta^*$. In particular, the linear term
$C_1q_\eta^*$ is proportional to
\begin{equation}
C_1=-2\,\frac{Im\,a_{\eta A}}{|a_{\eta A}|^2}\,.
\end{equation}

Information on the relative value of the imaginary part of the scattering length is of
special interest in the case of $\eta ^3$He interaction in view of existing discrepancy
between the theoretical values of $a_{\eta^3\!He}$ and the analysis of measurements of
$pd\to\eta ^3$He in Ref.\,\cite{Smyrski,Adam,Mersman}. According to the latter results
$Im\,a_{\eta ^3He}$ is about 9 times smaller than $Re\,a_{\eta ^3He}$, being in
disagreement with the theoretical predictions of \cite{Wycech3,Shevchenko2,FiAr3b}. Such
a strong suppression of the $\eta N$ inelasticity in a nucleus is also in contradiction
to the intuitive expectation that with increasing number of nucleons the inelastic
effects in $\eta A$ interaction should become more and more important, so that for
heavier nuclei the enhancement effect due to $\eta A$ attraction is completely
overshadowed by the absorption and is practically invisible \cite{Oset,Maghrbi}.

Also desirable are measurements of the spectrum in the reaction $\gamma
^4$He\,$\to\pi^0\eta^4$He in the region of low $\eta^4$He relative energies. As already
noted in Introduction, an enhancement effect due to $\eta$-nuclear attraction which is
rather well seen in the reactions $\gamma ^3$He\,$\to\eta^3$He and $pd\to\eta^3$He was
not observed in the case of $^4$He \cite{Frasc,Adlarson}. It is therefore important to
prove, whether also the peak in the distribution over the relative energy $\eta ^4$He in
the above reaction will disappear or at least will be much less pronounced in comparison
to that observed on $d$ and $^3$He.

\section{Conclusion}

We considered several aspects related to the coherent photoproduction of $\pi^0\eta$
pairs on the $s$-shell nuclei. As is discussed in Introduction, these reactions have some
clear advantages making them preferable to corresponding processes with hadrons as
probes. Because of relative weakness of the electromagnetic interaction, photo-induced
reactions are known to furnish a special opportunity to study effects of interaction in
the final state. Furthermore, the underlying elementary process $\gamma N\to\pi^0\eta N$
seems to be under control, in the sense that the results of different analyses
\cite{Doring,Horn2,FKLO} of the existing data agree with each other. This is in contrast
to the reactions $pd\to\eta pd$ or $pd\to\eta ^3$He where the driving mechanism is still
not completely understood \cite{Germond,Tengblad,Uzikov}. Furthermore, due to smallness
of the pion mass it tends to escape the interaction region with high velocity, and the
major fraction of the production events correspond to low relative velocity between
$\eta$ and the recoiled nucleus. This allows a cleaner way to study $\eta$-nuclear
interaction, in comparison to $pd\to\eta pd$ where $pd$ interaction in the final state
should always strongly affect the interaction between $\eta$ and the deuteron. Therefore,
measurements of these reactions may be an additional important source of information on
$\eta N$ low-energy dynamics.

One of the innovations of the present paper is a study of dependence of the total cross
section on the spin and isotopic spin of the target. Since $\pi^0\eta$ photoproduction
seems to be dominated by the $D_{33}$ wave, among the $s$-shell nuclei the largest cross
section is predicted for $^4$He whereas for $^3$He it appears to be twice as small as for
the deuteron.

We analyzed the effects of final state interaction in the region of low $\eta A$
relative energies. In the simplest case when the pole in the amplitude is close to the zero
energy, measurement of the distribution over the relative $\eta A$ energy may be utilized to estimate the relative
value of the imaginary part of the scattering length $a_{\eta A}$ using a simple
expression for the linear term in the polynomial ansatz (\ref{polyn}). This information
is clearly important for our understanding of the role of inelasticity in $\eta A$ low
energy interaction.

\section*{Acknowledgment}
Valuable discussions with Christoph Hanhart are much appreciated. We also acknowledge
support from the RF Federal programm "Kadry"(contract
14.B37.21.0786), MSE Program 'Nauka' (contract 1.604.2011) and from the Ministry of
education and science of Russian Federation (project 16.740.11.0469).

\end{document}